\documentclass{rmf-d}
\usepackage{nopageno,rmfbib,multicol,times,epsf,amsmath,amssymb,cite,here}
\usepackage[latin1]{inputenc}
\usepackage[]{caption2}
\usepackage{graphicx}
\usepackage{hyperref}
\usepackage{lineno}
\usepackage{color}
\def\rmfcornisa{Research OR Education of Physics \hfill\rmf\ {\bf ??} (*?*) ???--???
\hfill MES? A\~NO?}

\def\rmfcintilla{{\it Rev.\ Mex.\ Fis.\/} {\bf ??} (*?*) (????) ???--???}
\clearpage \rmfcaptionstyle 
\newcommand{\req}[1]{(\ref{#1})}
\newcommand{\muF}{\mu_{F}}

\newcommand{\nn}{\nonumber}
\newcommand{\sla}{\hspace*{-0.10cm}/}
\def\phiDA{\phi}
\def\taub{\bar{\tau}}

\def\taub{\bar{\tau}}

\begin{document}
\title{   On wide-angle photo- and electroproduction of pions to twist-3 accuracy 
\vspace{-6pt}}
\author{ K. Passek-Kumeri\v{c}ki\footnote{Presented at HADRON2021.}     }
\address{Division of Theoretical Physics, Rudjer Bo\v{s}kovi\'{c} Institute,
HR-10002 Zagreb, Croatia }
\author{ }
\address{ }
\author{ }
\address{ }
\author{ }
\address{ }
\author{ }
\address{ }
\maketitle
\recibido{day month year}{day month year
\vspace{-12pt}}
\begin{abstract}
\vspace{1em}  
The wide-angle photo- and electroproduction of pions is investigated within 
the handbag mechanism in which the 
$\gamma^{(*)} N\to \pi N'$ amplitudes factorize
into subprocess amplitudes,$\gamma^{(*)} q\to \pi q'$,
and form factors representing $1/x$-moments of generalized parton distributions (GPDs). 
The subprocess is calculated to twist-3 accuracy taking into account 
both the 2- and 3-body Fock components of the pion. 
The cross-sections are compared to experiment and spin-effects are discussed.
\vspace{1em}
\end{abstract}
\keys{exclusive processes, meson production, generalized parton distributions, distribution amplitudes, higher twist}

\begin{multicols}{2}

\section{Introduction}

The hard exclusive processes are successfully
described by the handbag mechanism in which
only one quark from the incoming nucleon
and one from the outgoing nucleon participate 
in the hard subprocess while all other partons
are spectators.
In particular, it was applied to the Compton
scattering 
$\gamma^{(*)} N \rightarrow \gamma N$,
the simplest probe of the nucleon structure,
as well as, to the meson electroproduction 
$\gamma^{(*)} N \rightarrow M N'$.
The prerequisite of this approach is the existence
of at least one large scale in the process
enabling the use of the perturbative expansion
in the strong coupling constant,
as well as, the twist expansion.
Two kinematical regions
were extensively investigated in the past.
The deeply virtual (DV) region is characterized by
the large virtuality $Q^2$ of the incoming photon 
and a small momentum transfer $(-t)$ from the incoming
to the outgoing nucleon.
In the wide-angle (WA) region $(-t)$ is large, as well as
$(-u)$ and $s$, while $Q^2$ is smaller than $(-t)$
($Q^2= 0$ in the case of photoproduction).
The all order proofs of factorization
exist for DV Compton scattering (DVCS) \cite{Collins:1998be}
and DV meson production (DVMP) \cite{collins}.
The process amplitudes factorize in hard perturbatively
calculable subprocess amplitudes and 
generalized parton distributions (GPDs) 
which encapsulate the soft hadron-parton transitions, 
i.e., and the hadron structure.
In contrast, for WA processes the general factorization proofs 
are still missing
but it was shown 
that the factorization holds to next-to-leading order in the strong coupling
for WA Compton scaterring (WACS)
\cite{rad98,DFJK1} and 
to leading-order for WA meson production (WAMP) \cite{huang00}.
In these cases it was argued that in the symmetric frame where 
skewness is zero
the process amplitudes can be represented as a product of subprocess
amplitudes and form factors that represent $1/x$ moments of GPDs
at zero-skewness.

Both DVCS and  WACS were widely investigated in the last decades 
and the handbag factorization achieved a good description 
of the experimental data.
But for the DV pion production
the experimental data indicate the importance of the transversely 
polarized virtual photons
\cite{hermes,clas12,defurne,JeffersonLabHallA:2020dhq}
and the need to go beyond the leading twist-2 calculation
since only the longitudinally polarized photons contribute to the latter.
The twist-3 calculation with transversity 
GPDs 
was proposed 
and the calculation including 
twist-3 2-body pion Fock component 
achieved already the successful agreement with the data \cite{GK5}.
Similarly, 
the experimental data for photoproduction 
\cite{anderson76,zhu05,clas-pi0} show that the
the twist-2 contributions for WA pion production 
\cite{huang00} are not sufficient.
In contrast to DVMP,
it was found that the twist-3 contribution to pion photoproduction
vanishes in often used Windzura-Wilczek approximation, i.e., when just
twist-3 2-body pion Fock components are considered \cite{signatures}.
Both 2- and 3-body twist-3 Fock components of $\pi_0$
were considered in \cite{KPK18} and the results were successfully 
fitted to CLAS data \cite{clas-pi0}.
We report here on extension of this work to 
the twist-3 prediction 
for $P=\pi^{\pm}, \pi^{0}$ WA photo- and electroproduction 
\cite{KPK21}.

\section{Application of handbag mechanism}

The helicity amplitudes ${\cal M}^P_{\nu,\kappa',\mu,\kappa}$ for 
$\gamma^{(*)}(\mu) N(\kappa) \rightarrow P(\nu) N'(\kappa')$
process, i.e,  
the (electro)production of the pseudoscalar meson $P$,
in the WA angle region
can be expressed as
\allowdisplaybreaks
\begin{eqnarray}
{\cal M}^P_{0+,\mu +}&=& \frac{e_0}{2}\, \sum_\lambda \Big[
             {\cal H}^P_{0\lambda,\mu\lambda}\, \left( R_V^{P}(t)
              +2\lambda  \, R_A^{P}(t) \right)
\nonumber\\ &&
       - 2\lambda \, \frac{\sqrt{-t}}{2m} \,
{\cal H}^P_{0-\lambda,\mu\lambda}\,\bar{S}^{P}_T(t)\Big]
\nonumber
\end{eqnarray} 
\begin{eqnarray}
{\cal M}^P_{0-,\mu +} &=& \frac{e_0}{2}\, \sum_\lambda \Big[
                 \frac{\sqrt{-t}}{2m} \, {\cal H}^P_{0\lambda,\mu\lambda}\, R_T^{P}(t)
\nonumber  \\
           && -2\lambda \, \frac{t}{2m^2} \, {\cal H}^P_{0-\lambda,\mu\lambda}\,S^{P}_S(t)\Big]
\nonumber \\ & &
                              + e_0 {\cal H}^P_{0-,\mu +}\,S^{P}_T(t)
\, ,
\label{eq:Mamp}
\end{eqnarray} 
where $m$ is a nucleon mass.

The soft form factors, $R_i^{P}$ and $S_i^{P}$ ($F_i^P$), represent specific flavor combinations 
of $1/x$-moments of zero-skewness GPDs ($K_i^{a}$) 
\begin{equation}
F_i^a(t) = \int_0^1 \frac{dx}{x}\, K_i^{a}(x,t) 
\,,
\label{eq:flavor-FF}
\end{equation}
where $x$ is the average momentum fraction 
of an active quark and $a$ is its flavour.
The form factors $R_V^P$, $R_A^P$ and $R_T^P$ are related to the helicity non-flip GPDs
$H$, $\widetilde{H}$ and $E$, respectively. 
The $S$-type form factors, $S^P_T$, $\bar{S}_T^P$ and $S^P_S$ are
related to the helicity-flip or transversity GPDs $H_T$, $\bar{E}_T$
and $\widetilde{H}_T$, respectively%
\footnote{The GPDs $\tilde{E}$ and $\tilde{E}_T$ and their
associated form factors decouple at zero skewness.}
.

\begin{figure}[H]
\centering
  \includegraphics[width=0.8\columnwidth]{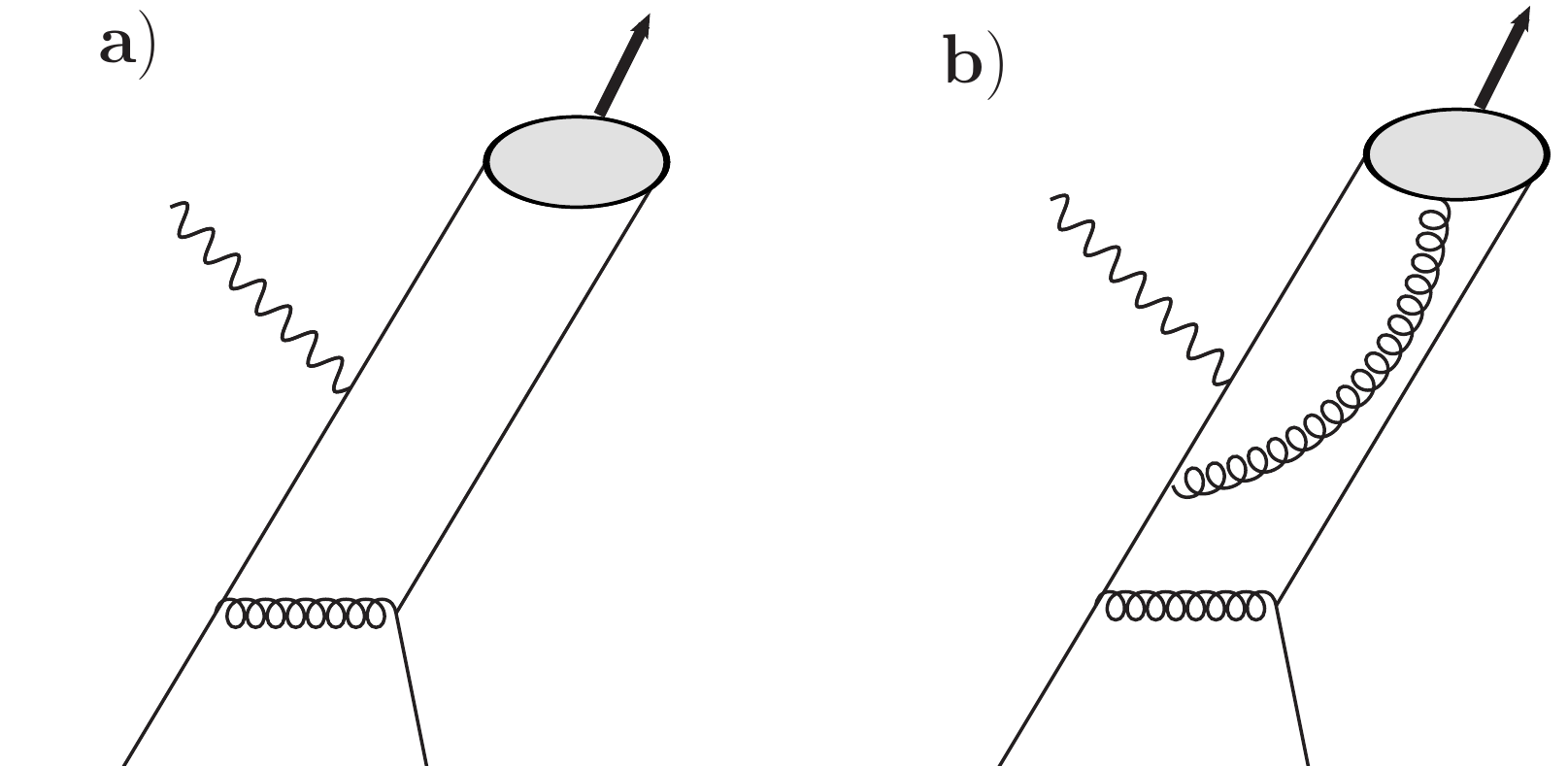} 
\caption{Generic diagrams corresponding to the 2- and 3-body subprocess amplitudes.}
\label{fig:H}
\end{figure}

The amplitudes
${\cal H}_{\mu,\lambda,\nu,\lambda'}$ 
correspond to the subprocesses
$\gamma^{(*)}(\mu) q(\lambda) \rightarrow P(\nu) q' (\lambda')$.
and they are calculated using handbag diagrams
as the ones
depicted on Fig. \ref{fig:H}.
The meson $P$ with momentum $q'$ is replaced by an appropriate
2- or 3-body Fock state.
The 2-body projector $\pi \rightarrow q\bar{q}$ 
\cite{beneke}
given by
\begin{eqnarray}
{\cal P}_{2}^{P}&\sim&f_\pi
   \,\Big\{\gamma_5 \,q\sla'\phiDA_\pi(\tau,\muF) 
 \nn \\ 
&&
                   + \mu_\pi(\muF) \Big[\gamma_5\,\phiDA_{\pi p}(\tau,\muF)  
 \nn \\ 
&&
- \frac{i}{6} \, \gamma_5 \, \sigma_{\mu\nu}\,\frac{q'{}^\mu n^\nu}{q'\cdot n}
                            \,\phiDA'_{\pi\sigma}(\tau,\muF)
  \nn \\ 
&&
                + \frac{i}{6}\, \gamma_5 \, \sigma_{\mu\nu}\,q'{}^\mu\phiDA_{\pi\sigma}(\tau,\muF) 
                           \frac{\partial}{\partial k_{\perp\nu}} \Big] \Big\}_{k_\perp\to 0}
\, ,
\label{eq:2proj}
\end{eqnarray}
contributes to the subprocess amplitudes
corresponding to the diagrams 
depicted on Fig. \ref{fig:H}a. 
The quark (antiquark) longitudinal momentum fraction is denoted by $\tau$ ($\bar{\tau}$),
and $k_T$ represents its intrinsic transverse momentum.
The first term in \req{eq:2proj} corresponds to the twist-2 part,
while the twist-3 part is proportional to the chiral condensate
$\mu_\pi=m_\pi^2 /(m_u+m_d)\cong 2$ GeV (at the factorization scale $\muF=2$ GeV).
This parameter is large and although the twist-3 
cross section for pion electroproduction is 
suppressed by $\mu_\pi^2/Q^2$ 
as compared to twist-2 cross section, 
for the range of $Q^2$ accessible in current
experiments the suppression factor is of order unity%
\footnote{
Twist-3 effects can also be generated by twist-3 GPDs. 
However, these are expected to be small and therefore
neglected.}%
.
The 3-body $\pi \rightarrow q \bar{q} g$ projector 
\cite{KPK18}
\\
\begin{equation} 
{\cal P}_{3}^{P} \sim
             f_{3\pi}(\muF) 
            \, \frac{i}{g} \, \gamma_5 \, \sigma_{\mu\nu} q'{}^{\mu} g_\perp^{\nu\rho}\, 
  \frac{\phiDA_{3\pi}(\tau_a,\tau_b,\tau_g, \muF)}{\tau_g} 
\label{eq:3proj}
\end{equation} 
contributes to the subprocesses amplitudes corresponding to 
Fig. \ref{fig:H}b.
The pion constituents are taken to be collinear to the pion
and $\tau_i$ denote their longitudinal momentum fractions.
The helicity non-flip amplitudes ${\cal H}_{0 \lambda,\mu,\lambda'}$ 
are generated by twist-2, while
the helicity flip ones ${\cal H}_{0 -\lambda,\mu,\lambda'}$ 
are of twist-3 origin.

Similarly to GPDs, the pion distribution amplitudes (DAs) $\phiDA$ 
encode the long distance effects 
and thus the pion structure.
In addition to twist-2 DA $\phiDA_\pi$ there are two 2-body twist-3 DAs,
$\phiDA_{\pi p}$ and $\phiDA_{\pi \sigma}$, and 3-body twist-3 DA 
$\phiDA_{3 \pi}$.
Twist-3 DAs are connected by equations of motion (EOMs)
\begin{eqnarray}
\tau \phi_{\pi p}(\tau) + \frac{\tau}{6}\, \phi'_{\pi\sigma}(\tau) - \frac1{3}\, \phi_{\pi \sigma}(\tau)
  & = & 
\phi^{EOM}_{\pi 2} (\taub) 
 \nn\\[0.2cm]
\bar{\tau} \phi_{\pi p}(\tau) - \frac{\taub}{6} \phi'_{\pi \sigma}(\tau) -\frac1{3} \phi_{\pi\sigma}(\tau)
& = & 
\phi^{EOM}_{\pi 2} (\tau)
\label{eq:EOMs}
\end{eqnarray}
where
\begin{equation*}
\phi^{EOM}_{\pi 2} (\tau) =
\displaystyle
 2 \frac{f_{3\pi}}{f_\pi\mu_\pi}\int_0^{\taub} \frac{d\tau_g}{\tau_g}\,
                                      \phi_{3\pi}(\tau,\taub-\tau_g,\tau_g)
\, .
\end{equation*}
Using EOMs and DA symmetry properties
the twist-3 subprocess amplitudes can be expressed
in terms of just two twist-3 DAs and 
2- and 3-body contributions can be combined.
Moreover,
EOMs lead to the inhomogeneous linear first order differential equation
which from known 3-body DA
$\phiDA_{3 \pi}$ 
\cite{braun90} 
fixes $\phi_{\pi p}$ (and $\phi_{\pi\sigma}$).
We note that 
in the derivation of  $q\bar{q}g$ projector and EOMs 
the same gauge has to be used 
for the constituent gluon, 
and we have consistently used the light-cone gauge.

For meson electroproduction 
both
transverse  
and longitudinal photons
contribute to twist-2 
subprocess amplitudes.
As expected in photoproduction limit ($Q^2=0$)
the longitudinal contribution vanishes,
while in DVMP limit ($t \to 0$) only 
the longitudinal photons contribute.
The twist-3 contributions both for transverse
and longitudinal photons have general structure
\begin{eqnarray}
\lefteqn{   {\cal H}^{P,tw3}}
\label{eq:Htw3}
\\ & = & {\cal H}^{P,tw3,q\bar{q}} +  {\cal H} ^{P, tw3,q\bar{q}g}
\nonumber \\
&=&
   \big({\cal H}^{P,\phi_{\pi p}} + \underbrace{{\cal H}^{P,\phi_{\pi 2}^{EOM}}\big)+
   \big( {\cal H}^{P,q\bar{q}g, C_F}} + {\cal H}^{P,q\bar{q}g, C_G} \big)
\nonumber \\
&=&
   {\cal H}^{P,\phi_{\pi p}} 
\; \;  + \quad \qquad {\cal H}^{P,\phi_{3\pi},C_F}
\qquad \quad + {\cal H}^{P,\phi_{3\pi},C_G}
\, .
\nonumber 
\end{eqnarray}
The twist-3 2-body contribution ${\cal H}^{P,tw3,q\bar{q}}$ 
is proportional to $C_F$ colour factor, while 
twist-3 3-body contribution has $C_F$ and $C_G$ proportional parts.
The $C_G$ part is separately gauge invariant,
while for $C_F$ contributions
only the sum of 2- and 3-body parts is gauge invariant 
(with respect to the choice of photon or virtual gluon gauge).
We have used EOMs to obtain that sum, as well as, the
complete twist-3 contribution expressed through just two
twist-3 DAs,  $\phi_{3\pi}$ and $\phi_{\pi p}$.
The twist-3 subprocess amplitude for longitudinal photons
vanishes both for $Q\to 0$ and $t\to 0$, i.e, 
for photoproduction and DVMP.
One finds that for 
photoproduction 
${\cal H}^{P,\phi_{\pi p}}=0$ 
\cite{KPK18}.
For DVMP ${\cal H}^{P,\phi_{\pi 2}^{EOM}}=0$, 
and although
for $t \ne 0$ one finds no end-point singularities,
in this limit one has to deal with
end-point singularities in ${\cal H}^{P,\phi_{\pi p}}$ 
\cite{GK5}.

\section{Numerical results}

For details on the form of the pion DAs, GPDs and parameters
used to obtain the numerical results 
as well as the
explicit expressions derived for the subprocess amplitudes
we refer to \cite{KPK21}.
The form factors related to both helicity non-flip
and transversity GPDs and only two independent pion DAs 
(twist-2 pion DA $\phiDA_\pi$ and twist-3 3-body pion DA $\phiDA_{3 \pi}$)
remain as soft physics input. 
The latter are accompanied by decay constants $f_{\pi}$ and $f_{3 \pi}$.

\begin{figure}[H]
\centering
  \includegraphics[width=0.7\columnwidth]{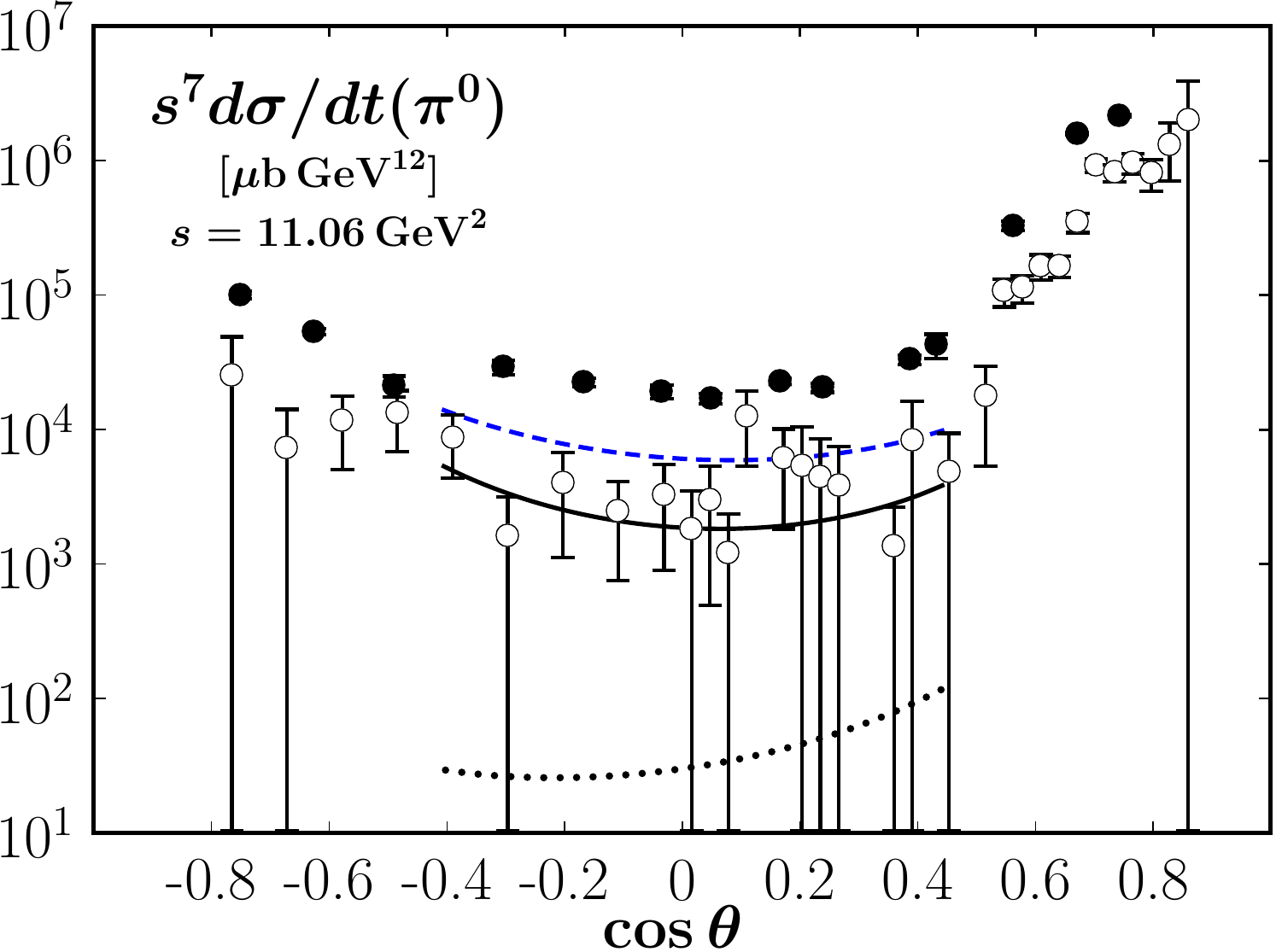} 
\caption{Photoproduction.}
 \caption{ The cross section for $\pi^0$ photoproduction.
    The solid (dotted) curve represents the full (twist 2) result.
    The dashed curve is obtained with the amplitudes taken at the fixed renormalization and factorization scale 
    $\mu_R=\mu_F=1\,$ GeV.
    Data taken CLAS \cite{clas-pi0} (open circles) and from SLAC \cite{anderson76} (the latter provided at $s=10.3$ GeV$^2$).}
\label{fig:photo}
\end{figure}

We comment here on few selected results.
In \cite{KPK18} the cross-section for $\pi^0$ 
photoproduction has been fitted to \cite{clas-pi0} data.
The results are displayed in Fig. \ref{fig:photo}.
As it was already noted in  \cite{huang00},
one can see from Fig. \ref{fig:photo} that twist-2 prediction 
lies well beyond the data. 
But by including the twist-3 contributions 
one obtains reasonable agreement with the experiment.
Twist-3 is more important in backward hemisphere ($\theta$ is c.m.s. scattering angle).
The twist-2 and twist-3 cross sections scale as $s^{-7}$ and $s^{-8}$, respectively.
Our results have effective $s^{-9}$ scaling which is a bit too strong.
As an attempt to modify this the fixed renormalization and factorization scales
were tested.
In \cite{KPK21} such an analysis was extended to $\pi^+$ and $\pi^-$
for which only few experimental data are available \cite{anderson76,zhu05}, 
The similar behaviour of photoproduction cross-sections was found. 

For pion electroproduction there are
four partial cross sections: 
$d \sigma_{L}/dt$,
$d \sigma_{T}/dt$,
$d \sigma_{LT}/dt$, and
$d \sigma_{TT}/dt$.
In \cite{KPK21} the theoretical predictions were given 
and importance of the measurement was stressed.
In the context of obtaining new information about transverse GPDs
we note that both for $\sigma_L$ and $\sigma_{LT}$ 
there is no twist-2 and twist-3 interference.
Given that the form factor $R_A^P$ contributing to twist-2 part 
is not unknown at large $(-t)$,
one could thus 
obtain the information 
on the $S_T$ form factor 
and gain knowledge on $H_T$ GPD. 
Furthermore,  $\sigma_{TT}$ that is suppressed for DVMP
could be used in WA limit to obtain the information on
$S_S$ and thus on completely unknown transversity GPD $\tilde{H}_T$.

\begin{figure}[H]
\centering
  \includegraphics[width=0.7\columnwidth]{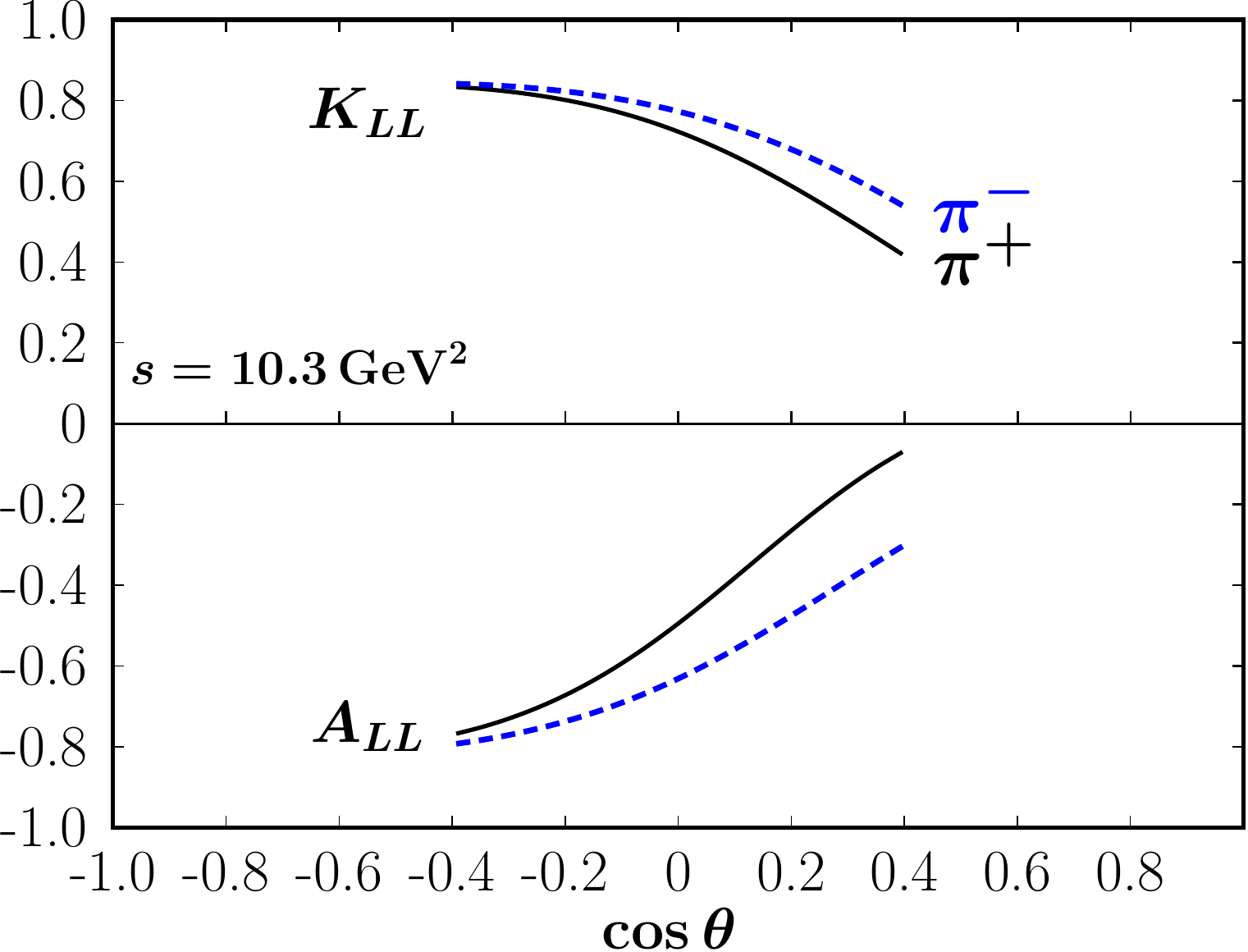} 
\caption{Results for the helicity correlation parameters $A_{LL}$ and $K_{LL}$ 
  for $\pi^+$ and $\pi^-$ photoproduction.}
\label{fig:AKLL}
\end{figure}

As often the spin-dependent observables offer additional insight 
which is less sensitive to particular parameters.
For meson photoproduction the most interesting are 
the correlations of the helicities of the photon and
that of either the incoming or the outgoing nucleon, i.e., 
$A_{LL}$ and $K_{LL}$, respectively.
It can be shown that
\begin{eqnarray}
A_{LL}^{P,tw2} &=&K_{LL}^{P,tw2}
\, , 
\nonumber 
\\[0.1cm]
A_{LL}^{P,tw3} &=&-K_{LL}^{P,tw3}
\end{eqnarray}
and so the measurement
of $A_{LL}$ and $K_{LL}$
offers characteristic signature for dominance
of twist-2 or twist-3. This is similar to the role that
the comparison of 
$\sigma_T$ and  $\sigma_L$ has in DVMP.
From Fig. \ref{fig:AKLL} it is clear that our 
numerical results suggest the dominance
of twist-3 for large $\theta$,
while twist-2 increases in the forward direction.

\section{Conclusions}

In this work we have summarized the recent application of the 
handbag factorization to WA photo- and electroproduction
of pions \cite{KPK21}.
In contrast to WACS, but like DVMP, the leading twist-2 analysis 
which involves helicity non-flip GPDs fails for WA photoproduction 
by order of magnitude.
We have obtained the twist-3 prediction for WAMP which includes
both 2- and 3-body twist-3 Fock components of the pions.
The $\pi^0$ photoproduction was fitted to the data \cite{KPK18}
Interesting helicity correlations
show that twist-3 dominates 
for 
$\pi^0$ channel, and mostly for $\pi^{\pm}$,
while in latter case twist-2 matters in forward hemisphere.
Different combinations of form factors along
with available data should allow
to extract the form factors
and to learn about large $-t$ behaviour of transversity GPDs
important for parton tomography.
The application to other pseudoscalar mesons \cite{KPK21a}
is straightforward.

{\it Acknowledgements}
This publication is supported
by the Croatian Science Foundation project IP-2019-04-9709,
by the EU Horizon 2020 research and innovation programme, STRONG-2020
project, under grant agreement No 824093
and by Deutsche
Forschungsgemeinschaft (DFG) through the Research
Unit FOR 2926, ''Next Generation pQCD for Hadron
Structure: Preparing for the EIC'', project number
40824754.

\end{multicols}

\medline

\begin{multicols}{2}

\end{multicols}
\end{document}